\documentclass[twocolumn,superscriptaddress,secnumarabic,amssymb, nobibnotes, aps, pre]{revtex4-1}

\usepackage{amsmath,bm,graphicx}
\graphicspath{{Figs/}}

\usepackage[usenames,dvipsnames]{color}
\definecolor{oceanboatblue}{rgb}{0.0, 0.47, 0.75}

\usepackage[colorlinks=true, citecolor=blue]{hyperref}

\usepackage[normalem]{ulem}

\begin{document}

\title{Living on the edge: Topology, electrostatics and disorder}%

\author{Tineke L. van den Berg}%
\email[Tineke L. van den Berg: ]{tineke\_vandenberg001@ehu.eus}
\affiliation{Centro de F\'isica de Materiales (CFM), 20018 Donostia -- San Sebasti\'an, Spain}
\affiliation{Donostia International Physics Center (DIPC), 20018 Donostia -- San Sebasti\'an, Spain}

\author{M. Reyes Calvo}%
\affiliation{CIC nanoGUNE, 20018 Donostia -- San Sebasti\'an, Spain}
\affiliation{IKERBASQUE, Basque Foundation for Science, 48013 Bilbao, Basque Country,  Spain}
\affiliation{Departamento de Fisica Aplicada, Universidad de Alicante, 03690 Alicante, Spain}

\author{Dario Bercioux}%
\email[Dario Bercioux: ]{dario.bercioux@dipc.org}
\affiliation{Donostia International Physics Center (DIPC), 20018 Donostia -- San Sebasti\'an, Spain}
\affiliation{IKERBASQUE, Basque Foundation for Science, 48013 Bilbao, Basque Country,  Spain}

\date{\today}%


\begin{abstract} 
We address the co-existence of massless and massive topological edge states at the interface between two materials with different topological phases. We modify the well known Bernevig-Hughes-Zhang model to introduce a smooth function describing the band inversion and the band bending due to electrostatic effects between the bulk of the quantum well and the vacuum. 
Within this minimal model we identify distinct parameter sets that can lead to the co-existence of the two types of edge states, and that determine their number and characteristics. We propose several experimental setups that could demonstrate their presence in two-dimensional topological systems, as well as ways to regulate or tune the contribution of the massive edge states to the conductance of associated electronic devices. Our results suggest that such states may also be present in novel two-dimensional Van der Waals topological materials.
\end{abstract}
\maketitle

\section{Introduction}

The study of edge states, or surface states for three-dimensional (3D) materials, goes back to the 1930s, when Tamm and Shockley studied bound states at the surface of a periodic lattice structure~\cite{Tamm_1932aa,Shockley_1939aa}.
The discovery of the quantized transversal conductance in the integer quantum Hall effect (IQHE) in the early 1980s~\cite{Klitzing_1980aa}, revealed the existence of a new type of edge state.  The measured quantized conductance is due to solely edge conduction, as it was realized only later~\cite{Laughlin_1981aa,Halperin_1982aa}. Contrary to classical edge states, the conducting edge states in the IQHE result from properties of the bulk of the system, namely its Landau levels | it turned out this was one of the first example of topological edge states~\cite{Thouless1982}. Not only are these states robust to disorder, but it is even desirable to invoke disorder for understanding why it is relatively easy to measure the conductance plateaus. Moreover, to capture the details of the spatial distribution and interactions between edge states, a proper description of the electrostatic potential at the edge is needed~\cite{chklovskii1992electrostatics,Guven_2003}. 


It was thought for a long time, one necessarily needed the breaking of time-reversal symmetry (TRS), in order to get quantized conduction at the sample edges. With the theoretical discovery of the topological insulator (TI), it was realized that different symmetry classes could also give rise to systems with edge conduction in absence of an external magnetic field, for reviews see~\cite{Hasan_2010aa, Qi_2011aa, Chiu_2016aa,Haldane_2017aa}. The first type of non-trivial system that was proposed was one giving rise to the quantum spin Hall effect (QSHE)~\cite{Kane_2005aa,Bernevig_2006ab}. In particular, this effect was predicted to be present in two-dimensional (2D)  mercury telluride quantum wells (QWs) grown on cadmium telluride substrates (HgTe/CdTe) ~\cite{Bernevig_2006aa}, for a recent review see Ref.~\cite{gusev2019mesoscopic}. Bandgap inversion in the bulk gives rise to two spin-locked, counter-propagating modes at the edges with linear dispersion (massless). These edge modes are protected from disorder by TRS, and the \(2e^2/h\) two-terminal conductance plateau extends throughout the entire gap. Here no electrostatics are invoked to understand their spatial distribution at the edge. The expected \(2e^2/h\) quantized conductance, was indeed measured about a year after the proposal~\cite{Koenig_2007aa}.
Quantum wells of HgTe/CdTe became one of the standard platforms for investigating the QSHE in 2D systems; nonlocal, dissipationless transport has been measured~\cite{Roth_2009aa}, the helicity of the edge channels was demonstrated~\cite{Brune_2012aa}, and the edge currents were visualized~\cite{Nowack_2013aa,ma2015unexpected}. 
Besides in 2D HgTe QWs, experimental signatures of edge states in the QSH regime have also been predicetd and observed  in other platform as: GaAs/InSb QWs~\cite{Liu:2008,Knez_2011aa, Suzuki_2013aa, Li_2015aa} and in monolayer 1T' phases of transition metal dichalcogenide crystals of WTe$_2$ and WSe$_2$~\cite{Fei_2017aa, Wu_2018aa, Ugeda:2018aa,Tang_2017}, bismuthene \cite{Yang_bismuthene,Reis_bismuthene,drozdov_bismuthene} and other layered materials~\cite{jacutingaite_2019}. 

Still, observation of conductance quantization in the QSHE is never as precise as is observed in the IQHE, and even in small and clean samples conductance is always fluctuating substantially, up to a 10 or 20 percent of \(2e^2/h\). Besides being due to residual bulk or surface conduction, several other mechanisms have been proposed for explaining the origin of the these fluctuations~\cite{Maciejko_2009,Stroem_2010,Crepin_2012,Schmidt_2012,Vaeyrynen_2013aa,Koenig_2013aa,Essert_2015aa}. The most important source of fluctuations is thought to be disorder, or to be specific disordered charge puddles in topological quantum wells, for example HgTe/CdTe. Although QSH edge states are in principle protected from disorder, disordered charge puddles can be a possible source of decoherence. The electrons, while wondering around in the puddle, may undergo inelastic scattering, losing coherence and potentially scattering back into a reverse-direction edge state, which would result in lower than \(2 e^2/h\) edge conduction~\cite{Vaeyrynen_2013aa,Koenig_2013aa,Essert_2015aa}. In the following, we will focus on a mechanism that can induce fluctuations above \(2e^2/h\), namely the co-existence of additional, edge states. Such Shockley type edge states can arise from electrostatic interface effects, such as band pinning or band bending, and the presence or absence of a topological edge state is no requirement. In contrast to the linearly dispersive (massless) topological states, such states have a parabolic-like dispersion relation, so are said to be massive edge (ME) states. Besides electrostatic boundary effects, another possible origin of ME states at topological interfaces, discovered in 1985 by A. Volkov and B. Pankratov~\cite{Volkov_1985aa}, is the smooth (instead of abrupt) band inversion at the edge. States resulting from this effect are called Volkov-Pankratov (VP) states in the literature, and these states are of topological origin, in the sense that they are the result of the lifting of band inversion between a topological and a trivial material. These VP states always accompany a massless topologically protected state, although in the original paper by Volkov and Pankratov the proposed band inversion was associated with a gradient in the doping percentage of IV-VI semiconductors~\cite{Pankratov_2018}. Recently, VP states have caught attention again, when they were observed to reside in 3D strained HgTe systems~\cite{Inhofer_2017aa,Tchoumakov_2017aa,Mahler_2019aa}. 

Here we investigate the presence of ME states of topological and/or electrostatic origin, in 2D topological systems, within the Bernevig-Hughes-Zhang (BHZ) toy model. Within this model, we implement smooth band inversion and electrostatic edge effects, in order to investigate the behaviour of ME states in a generic setting. We also add disorder to the model, thereby demonstrating the different response to disorder of unprotected and topologically protected edge states. We propose four experiments that could detect ME states, and strategies for tuning their contributions to conductance. Those contributions could be tune off if undesired when only clean topological transport is aimed for, or used for  selective switching in multi-valued logic devices~\cite{Seo_2014}. 

In Sec.~\ref{sec_electrostatics} we will discuss the case of edge states due to electrostatics in normal systems, and how the situation changes in topologically non-trivial systems. In Sec.~\ref{sec_BHZ} we will start by introducing the BHZ model (Sec.~\ref{ssec_BareBHZ}), and how we adapt the model in order to include interface effects (Sec.~\ref{ssec_ExtendedBHZ}). We will study the spectral properties, both  charge and spin degrees of freedom (Sec.~\ref{ssec_spectral}). We will then add disorder to the model, before continuing to investigate the transport properties (Sec.~\ref{ssec_transport}). In Sec.~\ref{sec_experiments} we will propose four different experimental setups for detecting massive edge states in 2D devices.

\section{Electrostatics near device edges}
\label{sec_electrostatics}

The fact that many semiconductors have low densities of free carriers, results in long space-charge regions. This again leads to bending of the band structure at the interface, particularly when an electric field is applied to the system by means of a gate electrode in a field effect transitor (FET) geometry, in which case pinning of the Fermi level at the interface can substantially enhance the bending. Depending on the sign and magnitude of the band bending the interface layer is classified as depletion, accumulation or inversion layer~\cite{Moench_2001aa}. If one would like to know the precise spatial variation of the bands at the interface of a specific material, one should solve the Poisson equation, together with the Schr\"odinger equation if quantum effects are important, which is generally the case for strong band bending. These two equations should be solved self consistently, taking into account the specific details of the material, and imposing overall charge neutrality. Without going into such detail, we can take a look at a very general, simple case. If one wants to know the precise spatial variation \(V(y)\) of the band bending near the interface, one must solve Poisson's equation
%
%
\begin{align}
\frac{d^2 V(y)}{d y^2} = - \frac{\rho(y)}{\varepsilon_0 \varepsilon_m}\,,
\end{align}
%
%
where \(\rho(y)\) is the density of charge per unit area (or unit volume in 3D), and \(\varepsilon_{m(0)}\) is the dielectric constant of the material (vacuum). Because of the fact that the band structure does not remain flat near the interface, a quantum well-like potential landscapes appears, creating room for states close to the interface. These edge states carry a certain charge, which is equilibrated behind the interface over a certain distance, usually called the interface length. For moderately strong bending so that quantum effects can be neglected, this slab of material at the interface is called the depletion layer, and one can affirm the density of charge is approximately constant in space, \emph{i.e.} assuming $\rho(y)\propto e N_\text{D}$, where $N_\text{D}$ is the density of charges. These assumptions result in a band bending that is quadratic in space
%
%
\begin{align}
V(y) =  - \frac{e N_\text{D}}{\varepsilon_0 \varepsilon_m} (y - y_\text{int})^2
\end{align} 
%
%
where \(y_\text{int}\) is the interface length over which the bending occurs. This length scale depends on the dopant density of the material and on how strong the bending is. For strong enough bending, and depending on if the bending is up or down, edge states might appear either below the conduction band (CB) or above the valence band (VB)~\cite{Moench_2001aa}.

The solution to Poisson's equation at interfaces is known for many different settings, with more or less intricate solutions. However, in topological insulators there is a new ingredient, which is the presence of a metallic state near the interface, namely the topological edge state. This topological `metallic' edge state will introduce new screening effects, likely more pronounced once a gate potential is applied to the system. In order to calculate the exact spatial electron density, one has to solve the Poisson equation and the Schr\"odinger equation self consistently, given all of the systems ingredients. This is an arduous task, moreover because in many realistic systems 
the charge neutrality point of this state can be buried in the valence band~\cite{Skolasinski_2018aa}. Doing this calculation, for example for HgTe or InAs/GaSb QWs, goes beyond the scope of this work. However, in order to study the general physical aspects of ME states, it turns out the exact details of the interface function are of minor importance, as long as some general features are correctly taken into account~\cite{Tchoumakov_2017aa}.

\section{Edge states in the topological BHZ model}
\label{sec_BHZ}
The QSHE in HgTe/CdTe and InAs/GaSb QW systems are usually addressed within an eight band $k\cdot p$ model~\cite{Bernevig_2006ab,Liu:2008}. It was shown that for a layer of HgTe larger than a critical thickness the QW undergoes a topological phase transition resulting in spin-locked in-gap edge states~\cite{Bernevig_2006aa}.  A similarly topological phase transition was predicted in InAs/GaSb QW system induced by the effect of a electrostatic gate~\cite{Liu:2008}.
For addressing transport properties, however, this $k\cdot p$ model is not very convenient because it demands substantial numerical resources. The Bernevig-Hughes-Zhang (BHZ) model is a reduction of the \(k \cdot p\) theory which contains only the four bands closest to the Fermi energy, while keeping most important physical aspects. This model is both accurate and convenient for studying in-gap spectral and transport properties. It can be studied in the trivial regime, or in the topological regime, depending on the sign of the gap parameter~\cite{Bernevig_2006aa,Koenig_2008aa}. In the following, we summerize the main properties of the BHZ model. We then present a modification of it that allows for investigating the consequences of interface effects, namely the emergence of ME states.



\subsection{The bare BHZ model}\label{ssec_BareBHZ}

The standard four-band BHZ model in spin- and band-subspace reads:
%
%
\begin{align}\label{eq:s1}
\mathcal H & = \begin{pmatrix} h(\bm{k}) &0\\ 0&h^*(-\bm{k}) \end{pmatrix}\,,
\end{align}
%
%
with the spin-subblock Hamiltonians
%
%
\begin{align}\label{eq:s2}
h(\bm{k})=
\left(\begin{smallmatrix}
(M+C)-(B+D)\bm{k}^2 & A k_+ \\
A k_- & (C-M)+(B+D)\bm{k}^2
\end{smallmatrix}\right)
\end{align}
%
%
where $A$, $B$, $C$, $D$, and  $M$ are material dependent parameters, \( \bm k =\sqrt{k_x^2+k_y^2}\) and \(k_\pm=k_x\pm \text{i} k_y\). The Hamiltonian in \eqref{eq:s1} is expressed in the following basis:
%
%
\[
\{|e \uparrow\rangle,\, |h \uparrow\rangle,\, |e \downarrow  \rangle,\, |h \downarrow \rangle\}, 
\]
%
%
where $e$ refers to electrons in the CB and $h$ is the VB, whereas $\uparrow/\downarrow$ are the spin eigenstates along the $z$-direction.
This model gives rise to a topologically non-trivial system when the following condition is fulfilled $0<M/2B<2$~\cite{Koenig_2008aa}. The numerical results presented in the main text are obtained via the following set of parameters typical for HgTe/CdTe QWs: $A = 0.3654$~nm$\,$eV, $B=-0.686$~nm$^2 \,\,$eV, $D=-0.512$~nm$^2$~eV, and the gap parameter in the topological regime is \(M=-10\) meV, which is half the gap width. This parameter set gives rise to a band structure as that depicted in Fig.~\ref{f_BHZbands}(a). The gap is \(20\) meV (\(= 2 |M|\)), as ideally expected in HgTe QWs at zero temperature. In the gap, between the conduction and the valence bands, lie two topological modes with a linear dispersion. The crossing point lies just under the conduction band. In most HgTe structures, one would expect a flatter valence band around the \(\Gamma\) point, with a camelback shape, and we should underline again that the BHZ model is constructed for reflecting the in-gap properties of 2D materials~\cite{Bernevig_2006aa,Ortner2002,Skolasinski_2018aa}, and should therefore not be use it to study the bulk properties of a system. However, as we are interested here in studying the in-gap (edge) properties, the BHZ model provides a good framework. 
A natural length scale that emerges from this model is \(\xi=\hbar v_\text F /|M|\), where \(v_\text F \) is the Fermi velocity. In HgTe/CdTe QWs, with a bulk Fermi velocity of \(v_\text F =5 \times 10^{5}\)~m\,s\(^{-1}\) and \(M = -0.01\) eV, this gives \(\xi = 200\) nm. We will later use this length scale when deciding on the interface length of our system.

\subsection{Extended BHZ model: interface effects}
\label{ssec_ExtendedBHZ}
\begin{figure}[!t]
\includegraphics[width=0.29\linewidth]{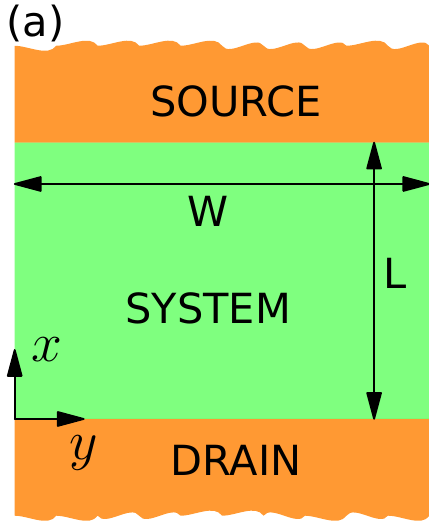}
\includegraphics[width=0.69\linewidth]{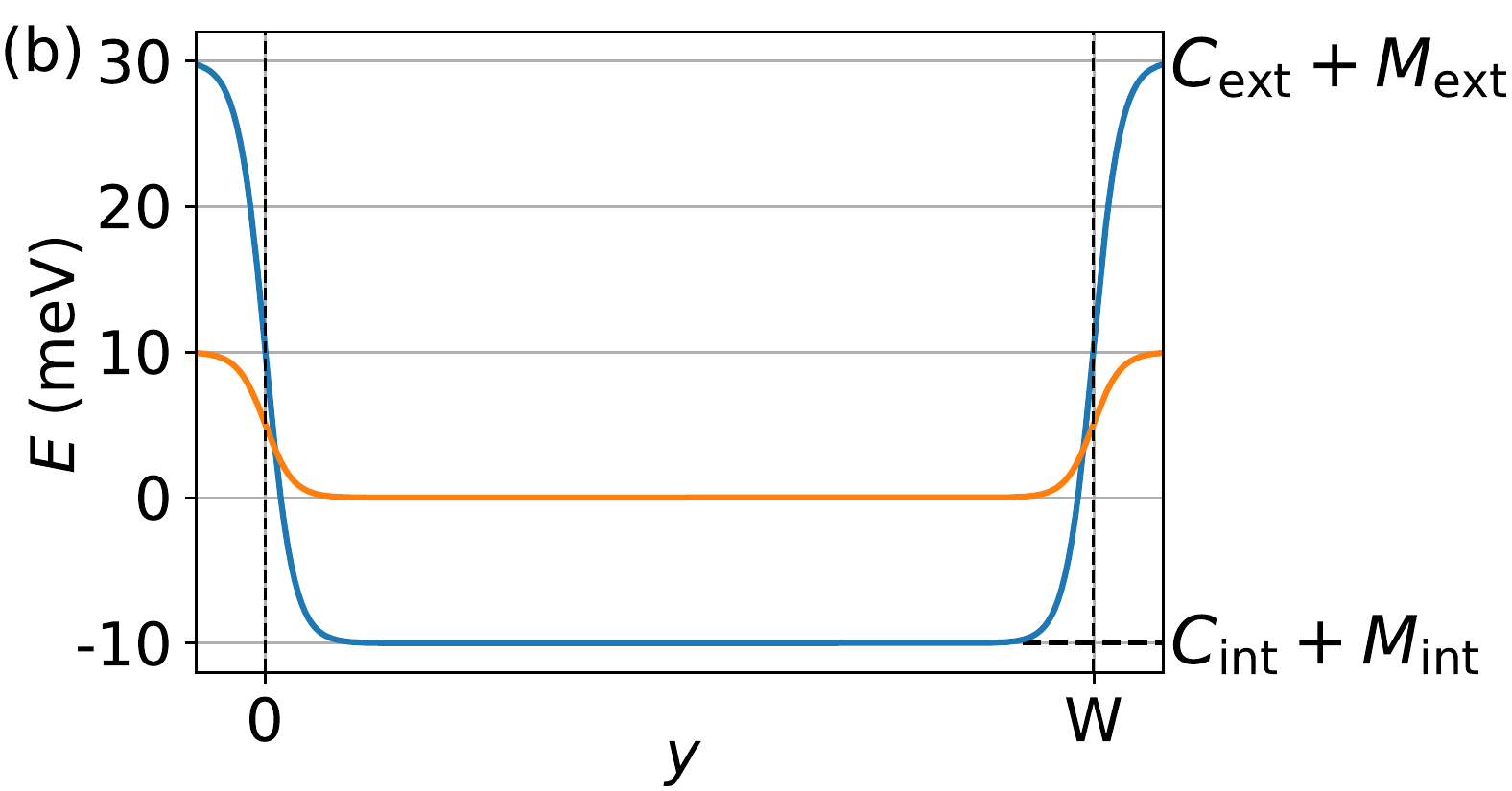}
\caption{\label{f_system} Sketch of the system (left), lateral structure of the Fermi energy (orange) and energy band onset (blue) with band bending for k=0. We here used the following parameters: \(C_\text{int}= ,\, C_\text{ext}=10\) meV, \(M_\text{int}=-10 \) meV and \(M_\text{ext}=20\) meV.}
\end{figure}

We will modify the bare BHZ model in order to make the band inversion and the electrostatic edge potential smooth and progressive. Therefore, in what follows, the parameters \(C\) and \(M\) will be functions of the lateral position \(y\), see Fig.~\ref{f_system}(b). 

%
%
%
%
Here, \(M(y)\) will account for the smooth inversion of the topological gap, and \(C(y)\) will account for band bending near the device edges. We will include these effects by writing the Fermi energy and the gap parameter of the system as a function of the form
%
%
\begin{align}
F (y) &= \frac{ F_\text{int} + F_\text{ext}}{2} \label{eq_Vy} \\
&+ \frac{F_\text{int} - F_\text{ext}}{2} \Big[ \tanh \left( \frac{y}{\ell } \right) -  \tanh \left( \frac{y-W}{\ell}  \right) -1 \Big] \,.\nonumber
\end{align}
%
%
where \(F_\text{int/ext} = M_\text{int/ext} \) (\(C_\text{int/ext} \)) is the gap parameter (Fermi level) inside/outside the system, respectively, and \(\ell\) is the interface length. The choice of the form of this function appears natural, fulfils all basic requirements and corresponds to a regular choice taken in the literature~\cite{Tchoumakov_2017aa,Moench_2001aa}. We will not justify it by means of self consistent calculations of the electrostatic landscape near the edge, which would be an extensive investigation by itself. However, changing this transition function to a different smooth function, such as \(F(y/\ell)\propto (y/\ell)^2\) or \(F(y) \propto \tanh \left[(y/\ell)^2\right]\), does not qualitatively change the results, which gives this choice a solid basis for this study.

In Ref.~\cite{Tchoumakov_2017aa}, Tchoumakov~\emph{et al.} analytically solved a similar model for a 3D TI with these type of boundary conditions. However, because the BHZ model is quadratic in momentum it can not be solved analytically with the same technique, and we therefore implemented it numerically using KWANT~\cite{Groth2014ia}.

\subsection{Spectral properties}
\label{ssec_spectral}
%
%
\begin{figure}[!t]
\begin{center}
\includegraphics[width=0.9\linewidth]{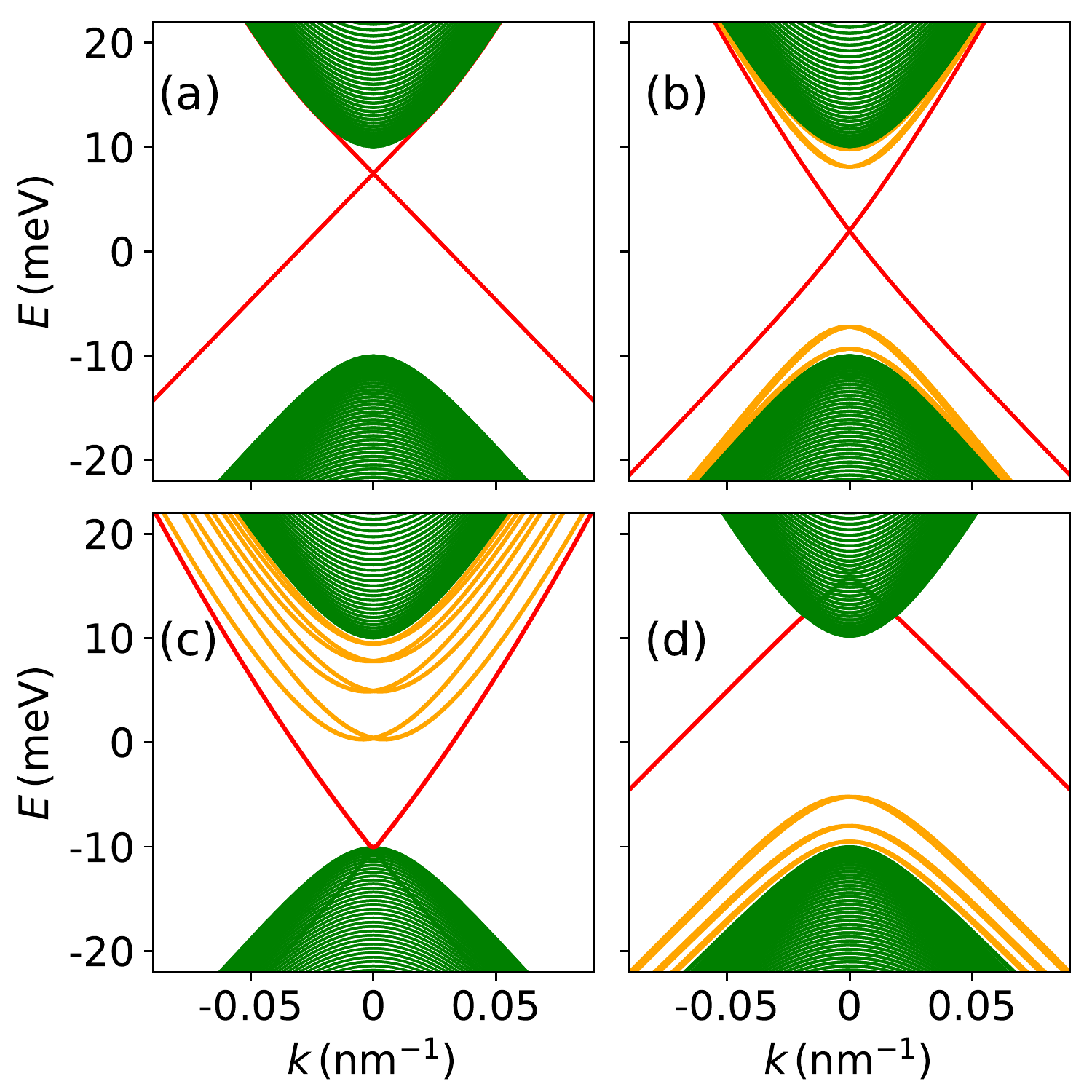}
\caption{\label{f_BHZbands} Bulk bands (green), VP states (yellow), topological edge states (red). (a) Bare BHZ model for HgTe parameters, with \(C(y) = C = 0\) and \(M(y) = M = -10\) meV. (b) With smooth band inversion with \(M_\text{int}=-10\) meV and  \(M_\text{ext} = 10\) meV with the interface length \(\ell = 200\) nm and \(C(y) = 0\). In (c) and (d) sharp band inversion \(M(y) = M  =-10\) meV as in the bare model, and Fermi pinning near the edges with \(C_\text{int} = -20\) meV and \(C_\text{ext} = 20\) meV in (c), \(C_\text{int} = 10\) meV and \(C_\text{ext} = - 10\) meV in (d), and \(\ell = 200 \)nm. }
\end{center}
\end{figure}
%
%

In order to obtain ME states in addition to the topological ones, one can change from an abrupt band inversion to a smooth band inversion, by taking \(M_\text{int}=-10\)~meV and \(M_\text{ext} = 10\)~meV and choosing an adequate interface length, which is done by taking \(\ell  \gtrsim \xi\) in order to have a smooth gap inversion. The longer the interface length is, the more ME states can be hosted near the interface. We here choose \(\ell = 200\)~nm, for which we get multiple, well detached ME states, which emerge both above the VB and under the CB, see Fig.~\ref{f_BHZbands}(b). In this situation the ME states are also called Volkov-Pankratov states, they are of topological origin and always appear accompanying a topological state. 

Another strategy for obtaining states at the edge, is applying a local potential near the edge, which would physically correspond to a band bending scenario. By adding a local edge potential to the outermost lattice sites in the BHZ system, the crossing point of the topological states can be moved up or down, as was already shown in Ref.~\cite{Skolasinski_2018aa}, and edge states may appear for strong enough potentials. Here we implement a physically more natural boundary potential given in Eq.~\eqref{eq_Vy}, which has the same effect of moving the crossing point down (or up).  This results in band structures as depicted in Fig.~\ref{f_BHZbands}(c) and~\ref{f_BHZbands}(d) if the sign of the potential is respectively negative or positive sign. In \ref{f_BHZbands}(c) the bending of the Fermi energy is positive, bending towards the conduction band,and reversely in \ref{f_BHZbands}(d), the Fermi energy bends towards the valence band. So depending on the sign of the bending, the ME states can hang under the CB \ref{f_BHZbands}(c), or lie above the VB \ref{f_BHZbands}(d). Similar results were also obtained for the case of 3D TI~\cite{Tchoumakov_2017aa}.  The band structure at the edge forms a sort of triangular confinement potential at the boundaries of the system; the ME states can be seen as states arising from this confinement, each of them presents a spin-split spectrum due to spin-orbit interactions, much the same as to what was predicted for quantum wires with spin-orbit interaction~\cite{Governale2002,Perroni2007,Smirnov2007,Bercioux2015}. 

For topological QWs exceeding a specific thickness, $k \cdot p$ calculations hint to a burying of the crossing point of the topological mode inside the camel back of the valence band~\cite{Ortner2002,Skolasinski_2018aa}. Within the BHZ model this configuration is recovered in the case of  Fig.~\ref{f_BHZbands}(c), and we will from now concentrate on this sort of configuration. The bending of the bands in this way, here implemented as a smooth onsite potential at the edge, accounts not only for finite size effects, but can also reproduce interface electrostatic effects, inhomogeneous gating, electrostatic screening due to the existence of topological metallic states, or any other interface effects resulting in a smooth modulation of the bands towards the edge. Independently of where it may come from, it is used here to obtain the physical situation of interest.

%
%
\begin{figure*}[!t]
\begin{center}
\includegraphics[width=\linewidth]{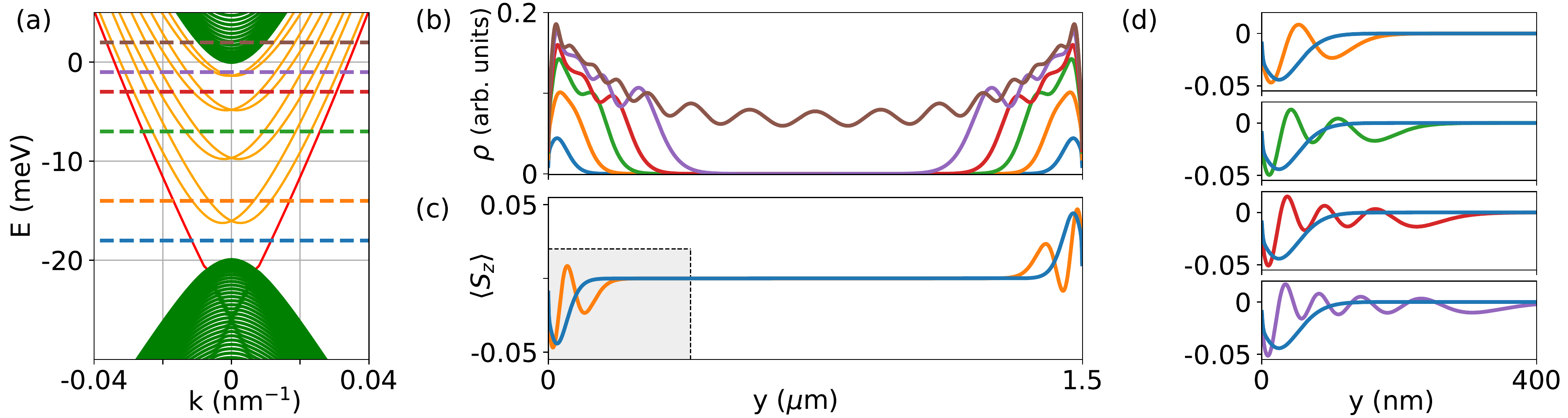}
\caption{\label{f_LDOS} Band structure with smooth band inversion and band pinning (a), together with the corresponding local density of states (b) at energies as indicated in the band structure. In blue one has the LDOS when the system contains only the topological edge states, one on each edge. For the orange-green-red-purple lines, one spin-split and double degenerate edge mode is added each time, hosting two edge states at each edge. The brown LDOS line is in the conduction band, giving rise to bulk states. (c) depicts the $z$-spin resolved local density of states, for the topological state (blue) and the first ME modes (orange). The grey area is shown in panel (d), containing the topological mode and consecutively adding the four ME states, as shown in the band structure of (a). For more details, see Appendix~[\ref{app_ldos}]. }
\end{center}
\end{figure*}
%
%
In addition to the band structure we can analyse the local density of state (LDOS), that we will refer to as \(\rho(x,y)\). In KWANT the LDOS is calculated for an open system, a scattering region connected to source and drain leads. The states in the scattering region, which is characterized by its scattering matrix, are the result of incoming modes at a given energy, from a given lead. In a clean system \(\rho\) is translationally invariant along the transport direction, so has only a lateral variation \(\rho(y)\), and it is sufficient to study line cuts, as in Fig.~\ref{f_LDOS}(b). Here we have summed over all modes, coming in from the top (source) lead, at energies specified in the band structure of Fig.~\ref{f_LDOS}(a). This reveals that the massive states observed in yellow in the band structure of Fig.~\ref{f_LDOS}(a),  only have a non-zero weight near the edges. It also tells us the topological states live closest to the edges, and the consecutive ME states spread out away from the edge. This spatial distribution of the LDOS should be observable in local probe experiments, such as scanning tunneling spectroscopy (STS) or scanning gate microscopy (SGM) experiments (c.f. Sec.~\ref{ssec_stm} and \ref{ssec_sgm}).
Furthermore, the ME states presents an oscillating behavior along the $y$ direction that is not observed for the topological one. The origin of the oscillations in the LDOS can be found in the interference of  modes at a fixed energy $ \mathcal{E}$, these are characterized by different values of the longitudinal momentum $k_x^i(\mathcal{E})$. We can express the LDOS of each mode of the system as
%
%
\begin{equation}
 \rho(y,\mathcal{E})=\sum_m \left|\sum_i \psi_m[y,k_x^i(\mathcal{E})] \right|^2\,,
\end{equation}
%
%
where $m$ is the associated mode index and we sum up over all the modes with energies smaller or equal to $\mathcal{E}$. Modes with a higher energy $\mathcal{E}$ oscillate more because they penetrate more inside the bulk region of the QW.

Using the same method as for the LDOS, we can calculate the spin polarization. As known from the QSHE, the electrons in the topological conduction channels are spin-\(z\) polarized, see Fig.~\ref{f_LDOS}(b, blue lines). The polarization is of opposite sign on opposite edges (upper panel), as expected. The massive edge states also show polarization with respect to spin-\(z\), which is asymmetric on opposite edges; similarly to the LDOS, the behavior is oscillatory. For a system containing the topological state and one massive state on each edge (orange curves), there are two positive oscillations, for a system containing two massive states at the edge (green lines) there are three positive oscillations, etc. The expectation values of the spin-\(x\) and spin-\(y\) components are zero, as expected. There is, in the BHZ model, no Rashba or Dresselhaus type of spin-orbit interaction, because the model does not include any structural inversion asymmetry. Spin-orbit interaction effects in a BHZ-type model were discussed in Ref.~\cite{Rothe_2010aa}.

However, one has to take into account the fact that realistic samples always contain some amount of disorder, and this might blur the effect somewhat, especially if one has to deal with charge puddles~\cite{gusev2019mesoscopic}. But as we will show in Sec.~\ref{ssec_stm}, even in the presence of disorder, in STS measurements one should still observe clear signatures of ME states.

\subsection{Transport properties}\label{ssec_transport}
%
%
\begin{figure*}
\includegraphics[width=0.6\linewidth]{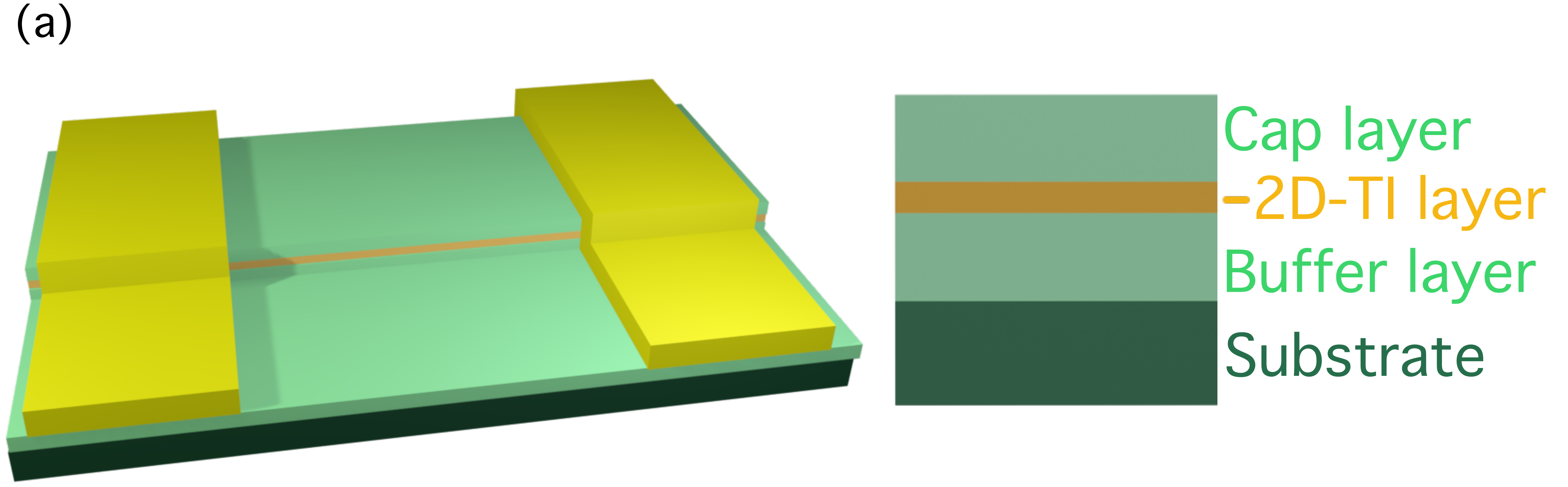}
\includegraphics[width=0.38\linewidth]{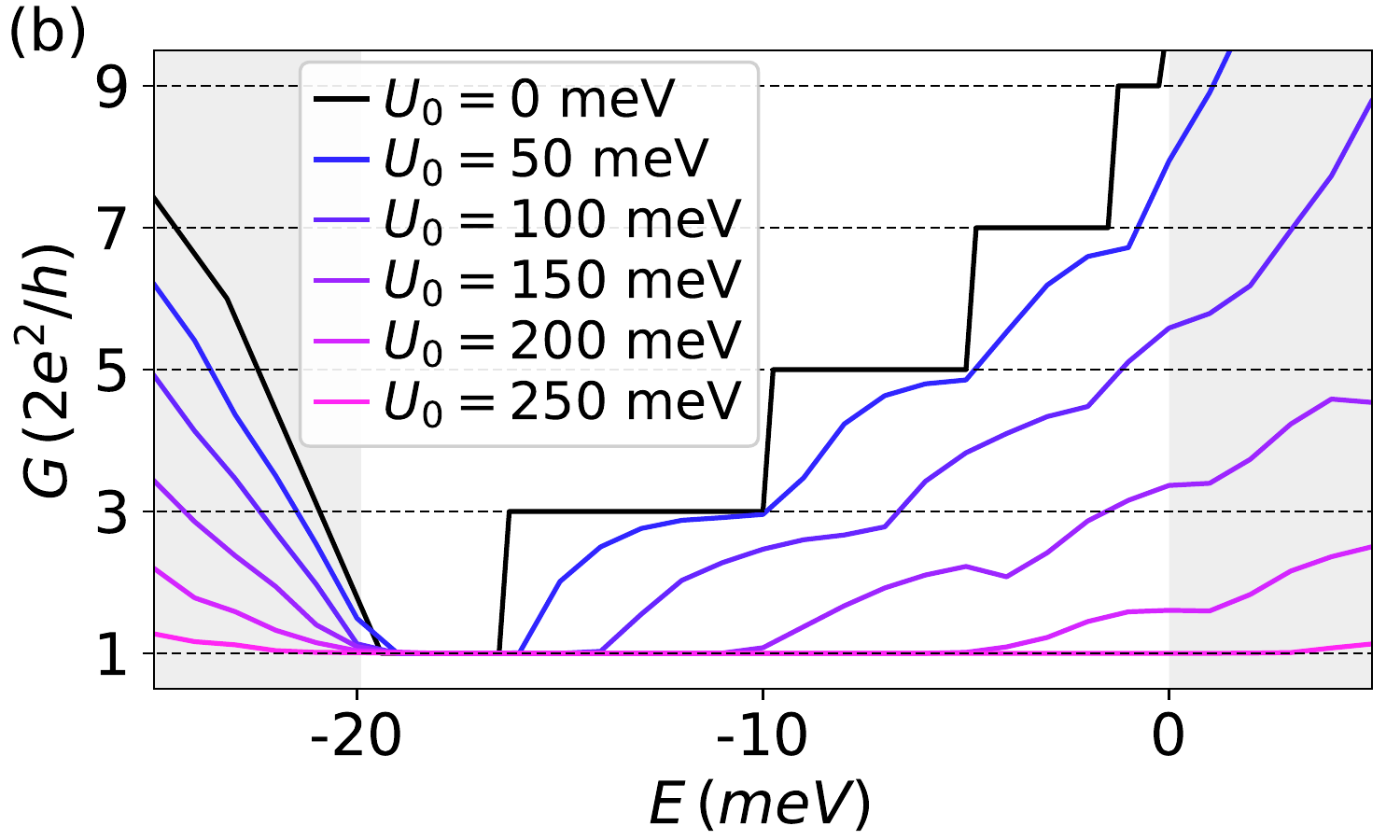}
\caption{\label{f_Gdisordered}  Conductance as a function of Fermi energy for \(500\times 1000\) nm systems, for a material with the band structure of Fig.~\ref{f_LDOS} (a). The black curve is for a clean system (no disorder), and the lines in purple are for successively stronger disorder, averaged over 80 disorder configurations. }
\end{figure*} 
%
%


As we mentioned before, the BHZ model is especially suited for doing transport calculations. We will start here with a device with a geometry such as depicted in Fig.~\ref{f_system}(a), \emph{i.e.} a scattering region connected to a source and a drain lead. The parameters of used in this paragraph are the same as in the previous paragraph, thus giving us the band structure of Fig.~\ref{f_LDOS}(a). By tuning the position of the Fermi level, simulating what is done experimentally by applying an overall back gate potential, one can scan through the entire band structure, thereby changing the number of ME bands crossing the Fermi energy and thus contributing to transport. For the system under investigation this means that moving the Fermi level from the VB, through the gap and into the CB, we successively add ME states, thus step-wise increasing the conductance.

Despite the fact that the additional edge states are not topologically protected, one can expect to observe their presence in transport measurements. If present, the two-terminal conductance in the gap will exceed \(2e^2/h\) for clean (or short) enough devices, see Fig.~\ref{f_Gdisordered}(b). In the case of a defect free sample, transport of ME states is ballistic and one observes a step each time the Fermi level crosses the energy onset of an ME state. Starting just above the VB with a \(2e^2/h\) conductance, one would add \(4 e^2/h\) at each opening of an ME state, as the two spin-split bands are doubly degenerate, and run on each edge of the sample.

In a more realistic scenario the presence of disorder will decrease the contribution of ME states to device conductance substantially. We implement Anderson type disorder, which means we take random on-site energies within an energy range \([-U_0/2, U_0/2]\). How much the contribution of the ME states decreases depends on both the device channel length and the disorder strength. Results for \(500\times1000\)~nm systems are shown in Fig.~\ref{f_Gdisordered}(b). In the weak disorder limit one should still observe steps in the conductance curve as one sweeps the Fermi level, such as for \(U_0=50-100\) meV. For systems with strong disorder, or for sufficiently long channels, the conductivity decreases to \(2e^2/h\) in the gap, meaning the only conducting state left in the system is the topologically protected one. Here we have considered a wide enough sample and uncorrelated disorder, so that percolation from one edge to the other across the bulk does not occur, even for strong disorder. Therefore, in our model, conduction can not decrease below \(2e^2/h\). In narrow devices, strong Anderson disorder can decrease the conductance below \(2e^2/h\), due to percolation between the two opposite conducting edges via favorable energy paths~\cite{Jiang_2009aa}. If one were to implement correlated disorder, charge puddles could form. Then the conductance could decrease below \(2e^2/h\) due to percolation between the lateral edges via charge puddles throughout the entire device, or due to trapping of particles in the puddles resulting in inelastic scattering.  

In Fig.~\ref{f_Gdisordered}(b) we observe a shifting of the energy of the conduction band opening as we increase the disorder strength. This is associated to a renormalization of the gap parameter and the Fermi energy due to the presence of the disorder, which was extensively discussed in Refs.~\cite{Groth_2009aa, Jiang_2009aa, Wu_2016aa} in the context of the topological Anderson insulator. The renormalization of the gap parameter \(M_\text{int} \rightarrow M_\text{int} + \delta M \) in the case of Anderson onsite energy disorder \(\propto \sigma_0\) is negative \(\delta M < 0\), thereby increasing the effective inverted band gap. 
 
While hints of the existence of ME states can be extracted from transport measurements, we discuss in the next section other experimental approaches that could provide a more direct observation and characterization of the properties of ME states. On the one hand, STS spatial maps directly address the distribution of the system LDOS, and can also be used to obtain the dispersion relations of the states through quasiparticle scattering, as shown in Sec.~\ref{ssec_stm}. On the other hand, by exploring the effect of a movable local gate in the transport measurements, we can distinguish between conduction from bulk or edge states by using a so-called SGM, as shown in Sec.~\ref{ssec_sgm}.

\section{Proposals for experimental detection and tuning}
\label{sec_experiments}


\subsection{Local density of states mapped by scanning tunneling spectroscopy}\label{ssec_stm}
%
%
\begin{figure*}[!t]
\begin{center}
\includegraphics[width=\linewidth]{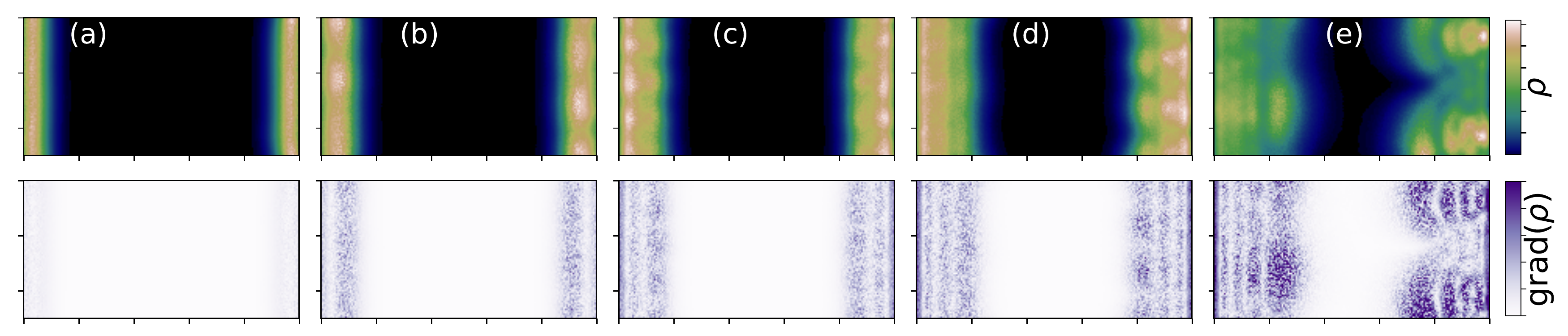}
\caption{\label{f_mapsLDOS} Local density of states within the gap for a system with disorder with \(U_0=20\) meV (upper panels), starting with only the topological state (left) and including ME states successively, at energies corresponding to the lines in Fig.~\ref{f_LDOS} (a). The gradient of the LDOS (lower panels), which can be calculated after measurement of the LDOS, reveals the pattern of the wavefunctions, as well as interference effects due to scattering on impurities.}
\end{center}
\end{figure*}
%
%
In the analysis of the local density of states in Sec.~\ref{ssec_spectral}, we have seen that it is possible to have several ME states in the system. The exact number will depend on the details of the system, namely on how strong the band bending is and how long the interface length is. The number of ME states that will be filled, depends on the position of the Fermi energy.  Experimentally, this can be set by the potential applied to an overall backgate electrode.

Assuming only elastic tunneling of electrons between the tip and sample, STS allows to directly map the LDOS of a sample. The applied bias voltage between tip and sample $V_\text{bias}$ determines the energy of electrons injected into the sample $eV_\text{bias}$ and the measured differential tunneling conductance at a given value of $V_\text{bias}$ is directly proportional to the LDOS at $E_\text{F}+eV_\text{bias}$~\cite{feenstrastm93}. Hence maps of differential conductance taken at a given $V_\text{bias}$ are effectively spatial maps of the LDOS, at a given energy $E_\text{F}+eV_\text{bias}$. One could also address the LDOS at different energies by mapping the LDOS at near-zero \(V_\text{bias}\), and use a back gate potential \(V_\text{BG}\) to tune the Fermi energy through the gap, producing maps at energies \(E_\text F (V_\text{BG})\).

In the example of certain HgTe QWs as in Fig.~\ref{f_LDOS}(a), moving from the VB, through the gap and into the CB, increases the number of ME states contributing to the LDOS. Those states spread out away from the edge as more and more are added. In order to demonstrate the presence of additional dispersive modes in a 2D TI, one can therefore monitor the changes of the spatial distribution of the LDOS as a function of energy via STS maps. A prerequisite for observing the ME states in transport experiments is that the system is not too dirty, as disorder will eventually localize the unprotected edge states. However, as can be seen in Fig.~\ref{f_mapsLDOS}, even in the presence of disorder the widening of the edge areas when increasing the number of available ME states can clearly be observed in the LDOS maps.

In Fig.~\ref{f_mapsLDOS} we remark the topological state is homogeneous in the \(x\)-direction, even in the presence of disorder. In contrast, the ME states display standing wave type interference patterns, for which the period depends on the energy at which the LDOS is probed. Measuring the relation between the energy and the interference period, one can extract the dispersion relation of the one quasi 1D edge states (for a related example in another quasi 1D system, see Ref.~\cite{sode_prb_grribbons,Buchs2018}). A parabolic dispersion relation from quantum interference measurements would be a smoking gun for the existence of ME states. Quantum interference experiments have been used to probe the electronic structure of other topological systems \cite{Nurit_review_2018}. At low energies, \emph{e.g.} in panels \ref{f_LDOS}(b)-\ref{f_LDOS}(d) the interference patterns are typical for quasi 1D modes, whereas in panel \ref{f_LDOS}(e) the interference pattern is more typical interference patterns for a 2DEG with impurities, as the highest ME state penetrates substantially into the bulk. 

However, performing STM experiments in buried structures is a rather challenging task. These phenomenology could be more easily explored in exposed 2DEG systems with inverted band structure, such as bismuthene~\cite{Reis_bismuthene,drozdov_bismuthene,Yang_bismuthene}, the single layers of the 1T' of some transition metal dichalcogenices~\cite{Tang_2017,Ugeda:2018aa} or other layered materials~\cite{jacutingaite_2019}. 
However, the necessary resolution in energy might be limited by the small size of the inverted gap, finite temperature effects and substrate interactions, which are yet more important for single layer materials.


\subsection{Conductance measurements with scanning gate perturbation}
\label{ssec_sgm}

%
\begin{figure*}[!t]
\begin{center}
\includegraphics[width=0.28\linewidth]{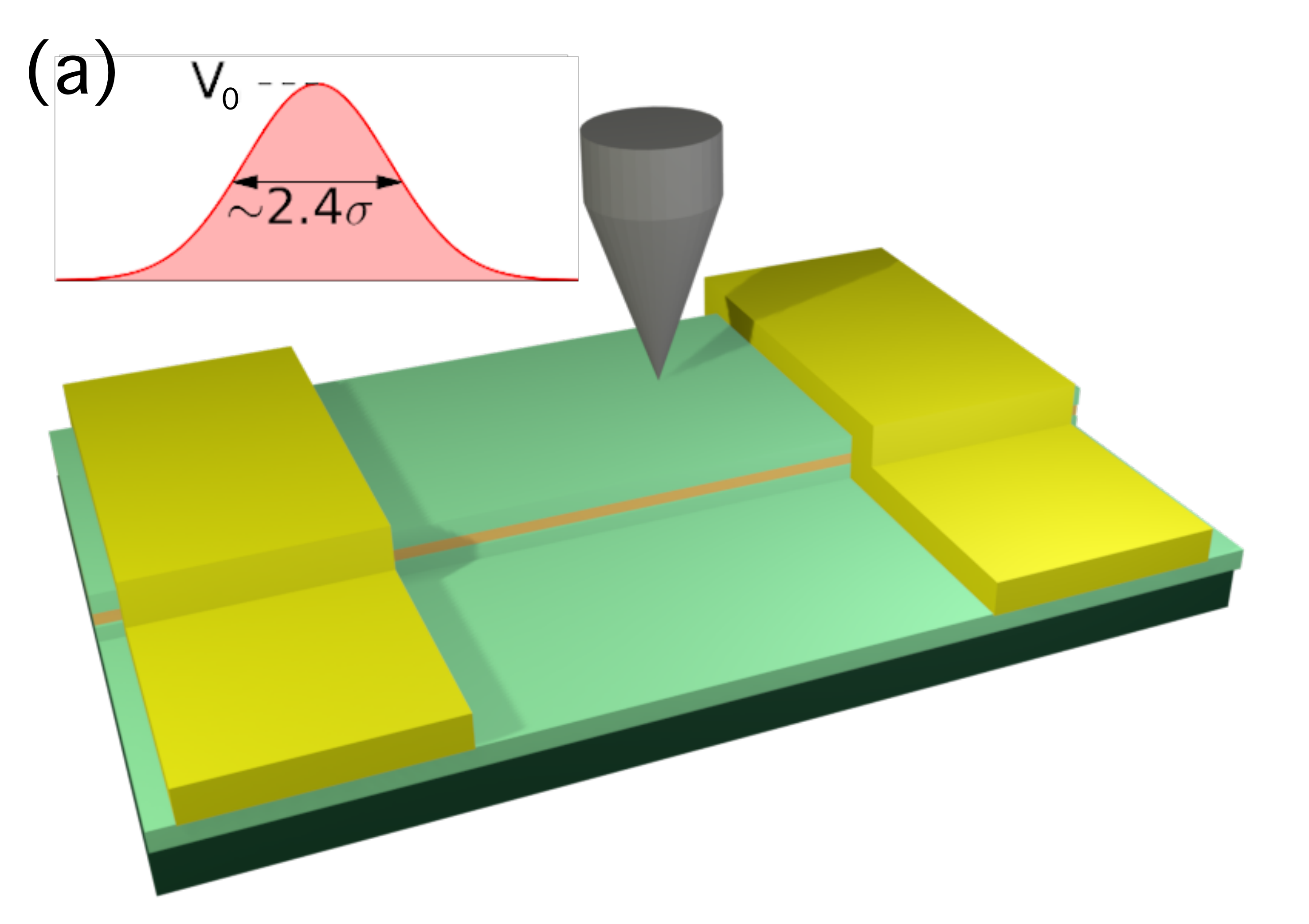}
\includegraphics[width=0.64\linewidth]{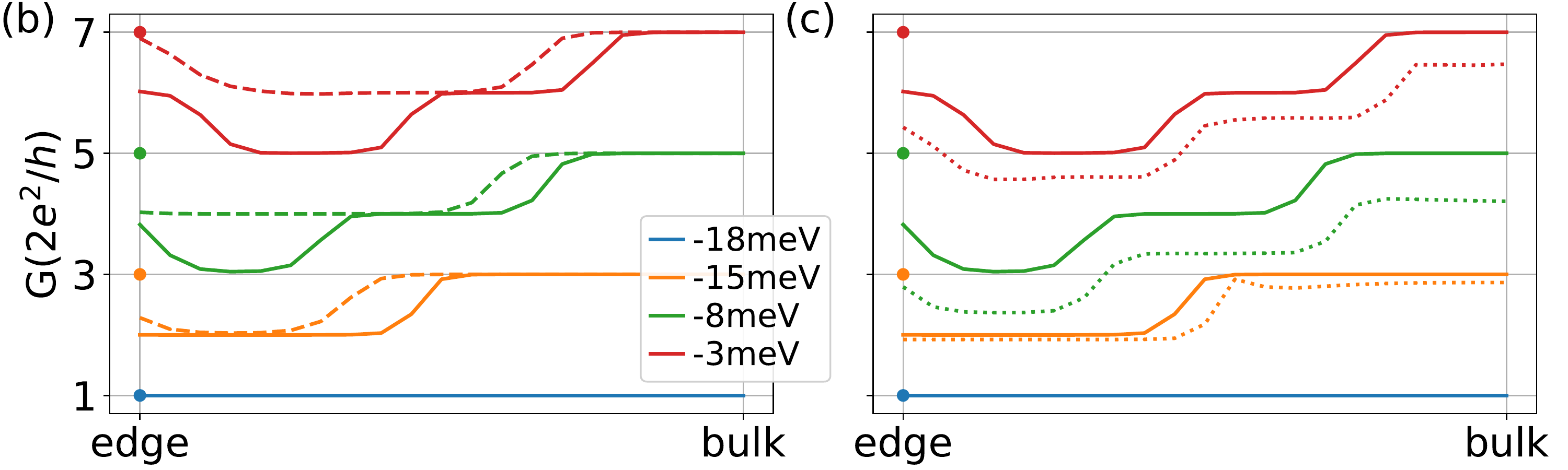}
\caption{\label{f_SGM} (a) Cartoon of the experimental SGM setup. (b) Conductance results in a clean system at four different back gates (\(-18\) meV to \(-3\) meV, blue to red | see corresponding horizontal cuts in Fig.~\ref{f_LDOS}(a) and tip energies of \(10\) meV (dashed) and \(20\) meV (solid line). (c) Same for a \(20\) meV tip energy compared to results for a weakly disordered system (dotted line). }
\end{center}
\end{figure*}
%
%
Scanning gate microscopy is a technique especially suited to spatially map scattering in quasi one-dimensional systems. In a SGM experiment we monitor the reaction of the system conductance to the perturbation induced by a local gate probe. The scanning gate is manoeuvred above the sample and a positive or negative voltage is applied to the tip, experienced by the system as a locally applied electric field of positive or negative sign. The tip is typically held rather far away from the sample, some 10-100~nm. Due to this distance and to the conical shape of the tip, this typically results in a rather large region of the sample being subject to the tails of the electric field distribution. The spatial resolution therefore depends on the intensity of the response of the conductance of the system to small changes on electric field. 

In the setup we proposed here, by perturbing the system with the tip, one locally shifts the energy level that sample area has within the band structure. Locally applying a negative electric field acts as placing a local barrier in the system, by lowering the energy of the electrons under the influence of the tip. On the other hand, a positive tip voltage will locally increase the energy of the underlying electrons.
If we assume that the Fermi energy of the unperturbed system lies within the bulk band gap of the topological 2DEG, the action of the tip can therefore open or close underlying edge modes, depending on the sign of the tip voltage. If the effective tip size is comparable to the system length, but smaller than the width, this will result in selective closing, or opening, of edge or bulk modes over the entire device length. If the effective tip size is small compared to the device system length, it will not result in addition conduction channels for positive tip voltages, as the new channels will not extend from one end of the device to the other. However, closing of channels can still occur, if the tip size is larger the the spatial (lateral) extend of the edge channel. For very small tip sizes, the system will see the tip action as an impurity, that will interfere with open modes, but will not cause the complete closing of channels.

In Fig.~\ref{f_SGM}, we simulate the perturbation effect on the sample of the electric field created by the tip in the form of a Gaussian on-site energy potential, see Fig.~\ref{f_SGM}(a). Taking into account the convention \(E= -e V\), where \(-e\) is the electron charge, the positive (negative) onsite energies represent negative (positive) tip voltages applied in experiment. The Gaussian potential of the tip is characterized by two parameters, its maximum height \(V_0\) and its half width \(\sigma\). If the tip is placed at position \((x_0,y_0)\) above the device, the onsite potential function is written as 
\begin{align}
V(x,y) = V_0 \, \mathrm{exp}\left[\frac{(x-x_0)^2+(y-y_0)^2}{2\sigma^2}\right]\,.\nonumber
\end{align}
The full width at half maximum (FWHM) is then \(2 \sqrt{2 \mathrm{ln}(2)} \sigma \approx 2.4 \sigma\) | Fig.~\ref{f_SGM}(a).

The numerical experiment is started by tuning the entire system so that its chemical potential lies within the gap, for example at \(E=-3\) meV as given by the red line in Figs.~\ref{f_LDOS}(a). The system can then be expected to have a conductance close to \(14 e^2 / h = (2+3\times 4) e^2 /h\), as there are three doubly degenerate ME states running on each edge. Then a suitable local electric field for the tip is chosen, which after testing turns out to be some tens of meV. This is the maximum height of the Gaussian on-site energy we use for simulating the tip, the half width is \(50\) nm for the simulations of Fig.~\ref{f_SGM}. The simulated tip is applied half way the length of the sample, at \(L/2\), and is run from the edge, up to the middle of the bulk, while calculating the resulting two terminal conductance. As a negative tip voltage suppresses states lying in the sample directly under it, conductance should decrease when the tip lies above open channels. Clearly (see Fig.~\ref{f_SGM}) such channels lie at the edge, at in-gap energies. In Fig.~\ref{f_SGM} is can seen that stronger tip potentials suppress more of the ME states. Also, even in a disordered system the features of edge state suppression should be readily observed.

On the other hand, applying a positive voltage to the local probe in our simulations, one locally increases the Fermi energy level of the underlying electrons. However, for small tip sizes this will not open a complete channel conducting from the source on one side, to the drain at the other side of the device. One can therefore not expect any increase in conductance. In experiments with real samples the signatures of scattering induced in topological states due to charge puddles has been shown using this technique~\cite{Koenig_2013aa} for devices with large stretches of edge. For shorter devices, we propose the selective tuning of states in a separate work~\cite{Priv_Comm_Person_PCA_Model}.

An alternative experiment to infer the spatial distribution of states is the characterization of the spatial distribution of supercurrents flowing in a Josephson junction with a 2D-TI material as weak link. For HgTe quantum wells, in Ref.~\cite{hart2014induced}, a certain evolution of the spatial distribution of current as a function of applied gate voltage can be observed, which could correspond to the tuning of massive states, but could also correspond to other inhomogeneous  electrostatic effects, or a combination of both. 

\subsection{Selective tuning edge conductance via dedicated  gate electrodes}
\label{ssec_edgeGate}

%
%
\begin{figure*}[t!]
\begin{center}
\includegraphics[width=0.28\linewidth]{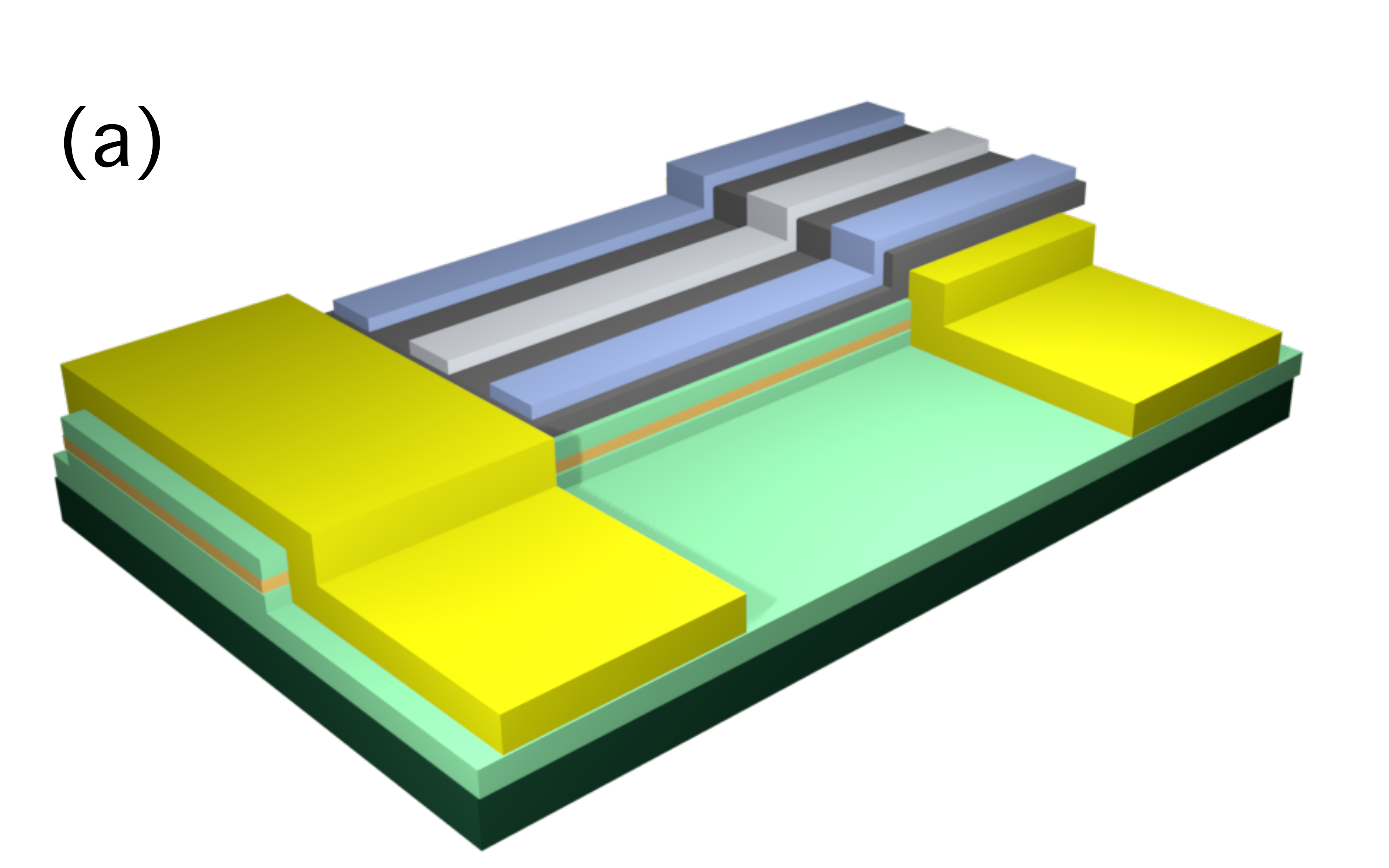}
\includegraphics[width=0.64\linewidth]{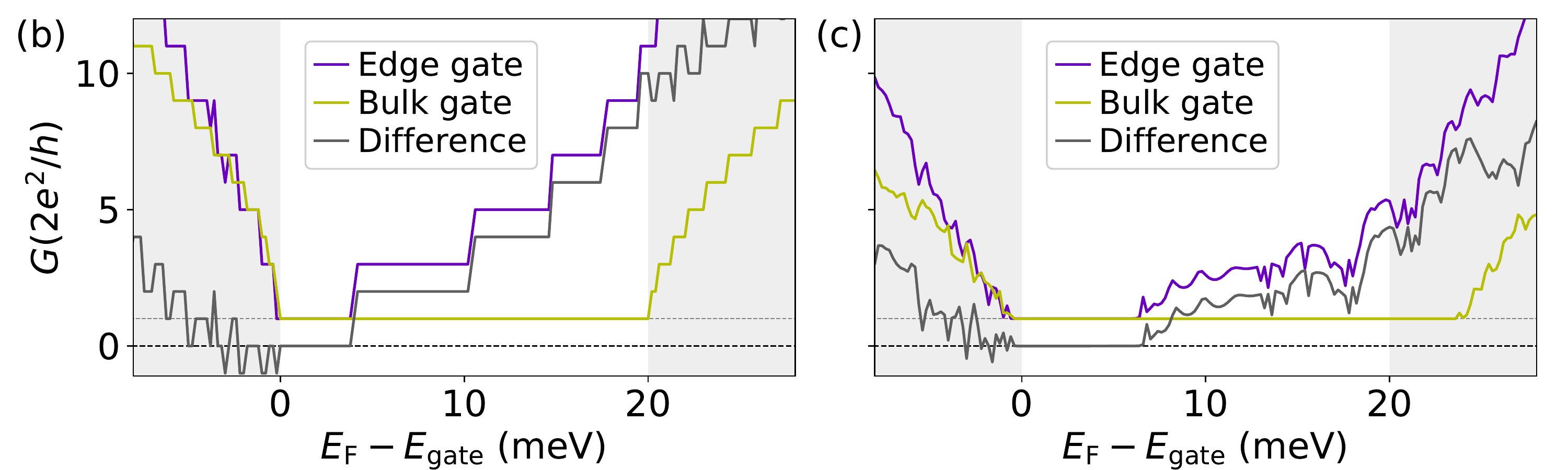}
\caption{\label{f_LocalGates} (a) Experimental setup with local edge and bulk gates. (b) Conductance for a system with a local edge gate and a local bulk gate, when scanning through the local edge gate or the local bulk gate. The difference between these two measurements clearly shows there is more than the topological edge state in the gap. (c) Also for a disordered system with \(U_0 = 100 \) meV. In the presence of this disorder the gap is larger due to the negative renormalization of the gap parameter.}
\end{center}
\end{figure*}
%
%
In this section we propose a experimental design that allows one to selectively tune the contributions to conductance of ME states. On the one hand, in the case of systems with strong disorder this setup will reveal the presence of additional in-gap ME states and unequivocally distinguish their contribution to conductance from bulk percolation. On the other hand, and more interestingly, this setup allows to isolate the contribution to conductance of the topological state from that of the ME extra states, which in most cases is undesired.
 
 The setup is unsophisticated and technologically feasible, which makes its realization easily possible. One needs a 2D QW layer, with a back gate for sweeping the entire sample through the gap, a split top-gate electrode for the edge and a top-gate electrode for the bulk, both running over the entire length | see Fig.~\ref{f_LocalGates}(a).

In our simulations we fixed the back gate so as to start from the top of the valence band, with a system containing only one topological state running at each lateral edge. Once this is fixed, one sweeps either the edge or the bulk gate through the energy range of the gap, while measuring the two-terminal conductance. 

In case there is anything more in the gap than just the topological state, one will find the conductance increasing in the gap while sweeping the edge gate. On the other hand, as there are no bulk states in the gap, the conductance should stay at (or close to) the \(2e^2/h\) value while sweeping the bulk gate. By subtracting one from the other, one can more precisely detect the difference. Even in systems with strong disorder, by subtracting bulk and edge gate sweeps, one can remove the disorder contribution to conductance.

This experimental setup would also allow for minimizing the effects of ME states. By applying edge gates to both lateral edges, and tuning those suitably, one could dispose of a substantial part of the interface effects as described above.


\subsection{Quantum capacitance measurements}
\label{ssec_capacitance}
\begin{figure}[t!]
\begin{center}
\includegraphics[width=0.45\linewidth]{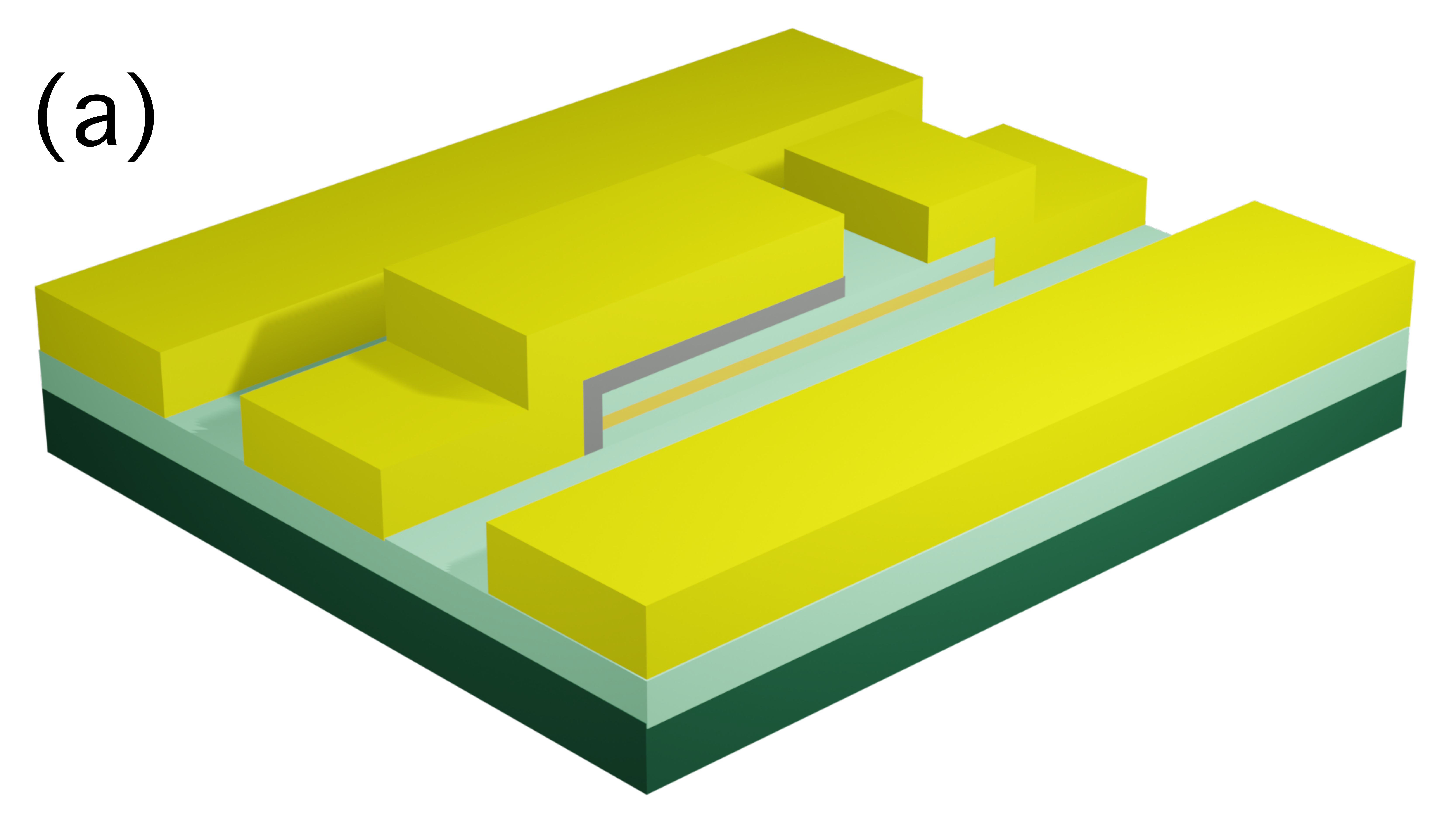}
\includegraphics[width=0.53\linewidth]{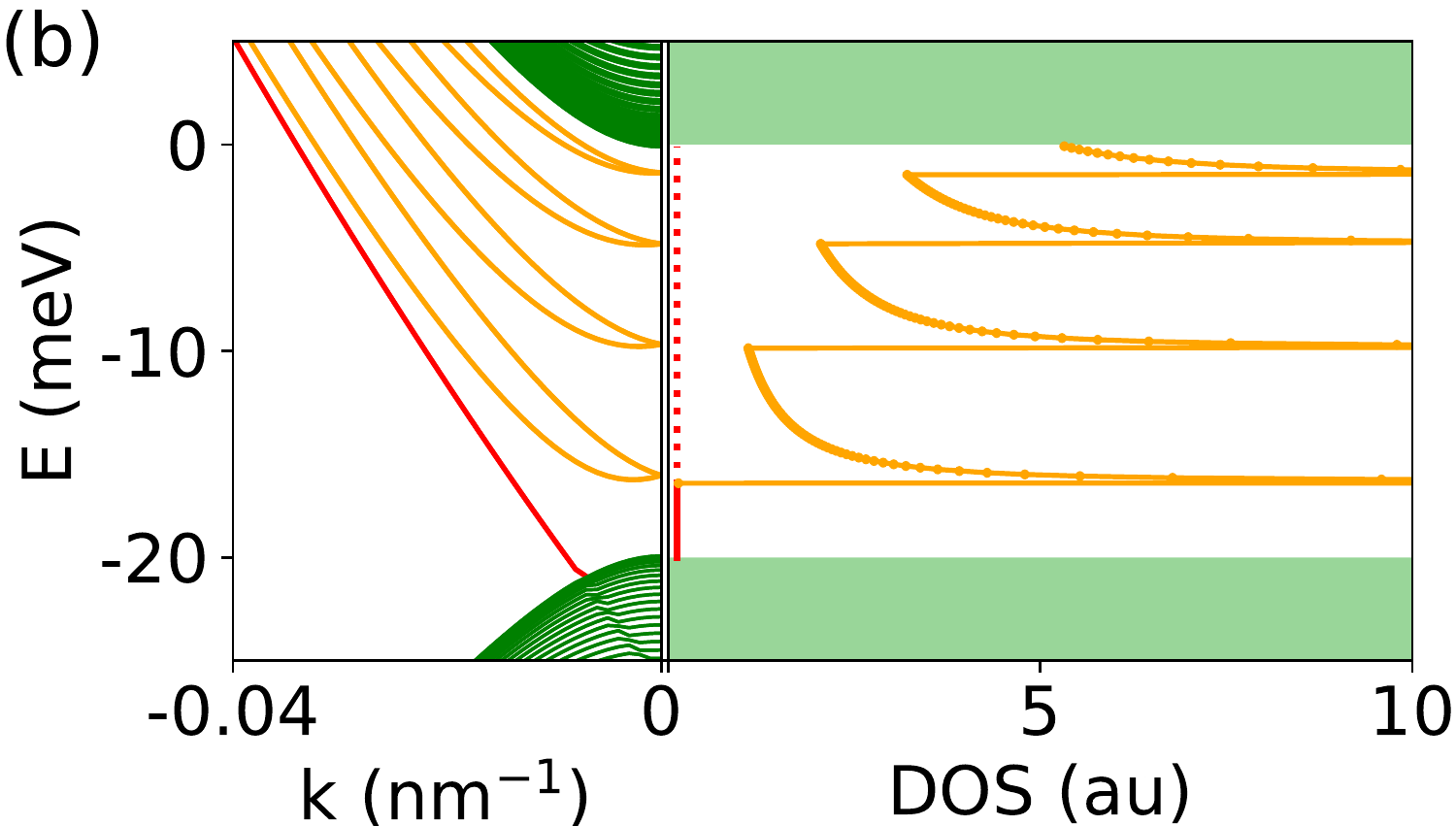}
\caption{\label{f_DOS} (a) Possible experimental setup, see Ref.~\cite{Dartiailh_2019aa}. (b) The in-gap density of states for a gap containing only topological bands (dashed red), and for a gap containing topological bands (solid red) and ME bands (dotted orange).}
\end{center}
\end{figure}

The capacitance of a device gives us information about how the electron density increases as we increase the electric potential applied to the system. It is the ratio of electronic charge to the applied electric potential. In nanoscale systems, besides the classical geometrical capacitance, there is also a contribution due to the quantum effects that become important both at small length scales and low temperature. The total capacitance is then given as
%
%
\begin{align}
    C^{-1}_\text t = C^{-1}_\text g + C^{-1}_\text q \,,
\end{align}
%
%
where \(C^{-1}_g\) is the geometrical capacitance and \(C^{-1}_q\) is the quantum capacitance. It can easily be shown that the quantum capacitance of a system is proportional to the density of states (DOS) around the electrochemical potential~\cite{Luryi_1988,Datta_2005aa}. Measuring quantum capacitance is therefore a common tool in studying electronic properties in the topological phase~\cite{Kozlov_2016aa,Kernreiter_2016aa,Braginsky_2019aa}. Experimental evidence of VP states in 3D TI where observed in quantum capacitance experiments~\cite{Inhofer_2017aa}, and recent experimental reports in 2D HgTe QWs show a values of the quantum capacitance exceeding the value due to the sole presence of topological edge states~\cite{Dartiailh_2019aa} (for a typical experimental setup see Fig.~\ref{f_DOS}(a). In the case of a low-temperature system, we can write express the quantum capacitance as
%
%
\begin{align}
    C_\text q &= - e^2 \int^{+\infty}_{-\infty} \text d E \, \rho (E) \left( \frac{\partial f}{\partial E} \right) = e^2 \rho (E_\text F + e V) \,,
\end{align}
%
%
where \( \rho (E_\text F + e V)\) is DOS at the Fermi energy shifted by an applied gate potential \(V\). The total capacitance is thus dominated by the smallest of the geometrical or quantum capacitance. In the case of gaped semiconductor the quantum capacitance in the gap is zero, as the in-gap density of states is zero. In topological materials, with an in-gap edge state, the DOS is constant, but non-zero in the gap | see Fig.~\ref{f_DOS}(b).

For a system of length \(L\) the 1D contribution of the topological state to the DOS is given by 
%
%
\begin{align}
    \rho_\text{QSH} (E) = \rho_\text{QSH} = \frac{2 L }{\pi \hbar v_\text F }\,,
\end{align}
%
%
giving a quantum capacitance of \(C_\text q ^0 = 0.19\) nF~m\(^{-1}\)~\cite{Entin_2017}. If there are in-gap ME states however, the DOS behaves very differently, having in-gap contributions of the ME states with parabolic-like dispersion. The DOS of these states is written as
%
%
\begin{align}
    \rho_\mathrm{ME} (E) = L \sum_n \frac{\sqrt{2m^*}}{\pi \hbar} \frac{\Theta (E-E_n)}{\sqrt{E-E_n}}\,,
\end{align}
%
%
where \(m^*\) is the effective mass of the charge carriers and \(E_n\) is the onset energy of the \(n\)th massive band~\cite{grosso2000}. This gives a quantum capacitance per unit length of 
%
%
\begin{align}
    C_\text q = C_\text q ^0 + \frac{e^2 \sqrt{2m^*}}{\pi \hbar} \sum_n \frac{\Theta (E-E_n)}{\sqrt{E-E_n}}\,,
\end{align}
%
%
implying that \(C_q > C_\text q ^0 \) in the presence of ME states. Here we are supposing the quantum contributions come only from (quasi) 1D edge states. In order to show this is indeed the case, one could measure the capacitance of systems of different length~\cite{Dartiailh_2019aa, Inhofer_2017aa, Fang_2007aa}. Doing this type of measurement, one can separate the geometric capacitance contribution from the quantum capacitance. One can also do AC microwave capacitance spectroscopy that additionally gives access to the resistive response of the system, reflecting on the ability of the system to conduct. This resistive part should therefore additionally give information about how many states contribute to transport~\cite{Pallecchi_2011aa}.
%
%
%
%

\section{Conclusion}

In this work, we have studied quantum wells hosting two-dimensional topological insulators within the Bernevig-Hughes-Zhang model. We have assumed that the mass and the onsite energy terms can vary smoothly at the interface between the bulk of the quantum well and the vacuum. We have shown the appearance of massive edge states in addition to the standard linearly dispersing mode of the quantum spin Hall effect. These massive edge states are characterized by a finite probability only close to the boundary of the system and by a spin-split parabolic-like energy dispersion. We have shown how these states can strongly affect the transport properties of a two-terminal system: the conductance of the system can increase above the nominal value of $2e^2/h$ of the topological states. However, due to the parabolic energy dispersion, these massive edge states are susceptible to the effect of local disorder. We have proven that in the case of strong disorder, their effect of the transport properties can be completely washed out. We have proposed various experimental setups that could pave the way to the detection and tunability of these massive edge states; these are mostly based on employing local probes and the design of local electrodes. Realistic samples are characterized by a more complex type of disorder as in-homogeneity and charge puddles. The latter can be included in our model, but their presence should not modify our findings substantially.
Our results are applying to the case of HgTe/CdTe and also to InAs/GaSb quantum wells; additionally, the general features we have shown should also be observable in two-dimensional materials presenting the quantum spin Hall effect as silicene, bismuthene and other van-der-Waals topological materials. 

\section*{Acknowledgements}
Discussions with E. Boquillon, A. De Martino, D. Goldhaber-Gordon, C. Gorini,  S. Tchoumakov, M. Wimmer are acknowledged. 

The work of TB and DB supported by the Spanish Ministerio de Econom\'ia y Competitividad (MINECO) through the project  FIS2014-55987-P,  by Spanish Ministerio de Ciecia, Innovation y Universidades (MICINN) through the project FIS2017-82804-P, and by the Transnational Common Laboratory \emph{Quantum-ChemPhys}. MRC acknowledges funding from the Spanish Government through project MAT2017-88377-C2-2-R and the Generalitat Valenciana through grant Cidegent2018004.

\begin{appendix}

\section{A deeper look into the local density of states}\label{app_ldos}
%
%
\begin{figure*}[!tb]
\begin{center}
\includegraphics[width=0.95\textwidth]{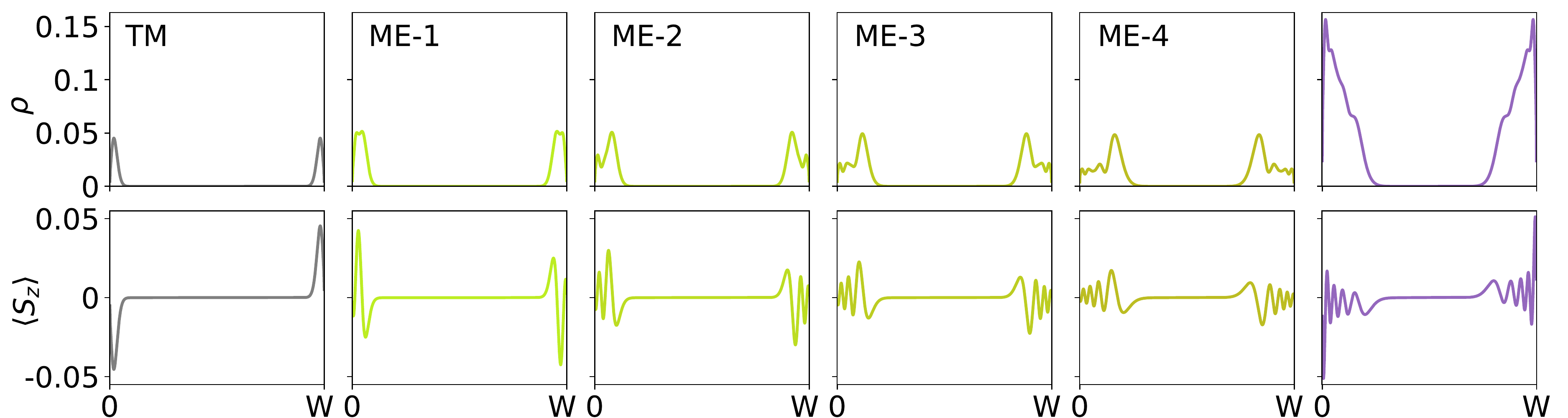}
\caption{\label{f_modes} Topological mode (left, grey), four different ME modes (middle, shades of green) and the sum of all available states at \(E=-1\) meV (right, purple), corresponding to the purple line in Fig.~\ref{f_LDOS}(a).}
\end{center}
\end{figure*}
%
%
Here we will discuss how the different modes of the wave function add up to the local density of states. For each mode \(m\), the wave function is the sum over all intersections \(k_x^i\) in the band structure at energy \(\varepsilon\),
%
%
\begin{align}
    \psi _m (y, \varepsilon) = \sum_i \psi _m [y, k_x^i(\varepsilon)]\,,
\end{align}
%
%
where \(y\) gives the lateral dependence of the wave function. The local density of states is the square of the wave function, summed over all modes available at the Fermi energy,
%
%
\begin{align}
    \rho (y, \varepsilon) = \sum_m \rho_m (y, \varepsilon) = \sum_m |\psi_m (y, \varepsilon) |^2
\end{align}
%
%
In similar fashion, for the spin polarisation, we have
%
%
\begin{align}
    \langle \sigma_j \rangle  (y, \varepsilon) &= \sum_m \langle \sigma_j  (y, \varepsilon) \rangle_m \nonumber \\
    &= \sum_m \sum_i \psi _m^* [y, k_x^i(\varepsilon)] \sigma_j \psi _m [y, k_x^i(\varepsilon)]
\end{align}
%
%
As the ME states are two doubly degenerate spin-split modes, their spin components have both positive and negative values near each edge, resulting in local oscillations of each mode. Consecutive modes of higher energies move more and more into the bulk, resulting in more oscillations for higher energy modes, as can be seen from Fig.~\ref{f_modes}. We also observe that for each consecutive mode, the main contribution to the \(\rho(\varepsilon)\) moves farther away from the edge. This results in an overall oscillating \(\rho(\varepsilon)\), with the number of oscillations depending on the number of ME modes available at energy \(\varepsilon\).

\end{appendix}

\bibliography{MassiveEdgeStates}

\end{document}